\documentclass[%
 reprint,
 amsmath,amssymb,
 aps,
 prl,
 ]{revtex4-2}

\usepackage{comment}

\usepackage[utf8]{inputenc}

\usepackage{amsmath,amsfonts,amssymb}
\usepackage{mathrsfs}

\usepackage{amsthm}

\usepackage{physics}

\usepackage[mathcal]{euscript}

\usepackage[dvipdfmx]{graphicx}
\usepackage{url}
\usepackage{bm}

\usepackage{color}

\usepackage{here}
\usepackage{braket}

\usepackage{xcolor}

\DeclareFontFamily{U}{BOONDOX-calo}{\skewchar\font=45 }
\DeclareFontShape{U}{BOONDOX-calo}{m}{n}{
  <-> s*[1.05] BOONDOX-r-calo}{}
\DeclareFontShape{U}{BOONDOX-calo}{b}{n}{
  <-> s*[1.05] BOONDOX-b-calo}{}
\DeclareMathAlphabet{\mathcalboondox}{U}{BOONDOX-calo}{m}{n}
\SetMathAlphabet{\mathcalboondox}{bold}{U}{BOONDOX-calo}{b}{n}
\DeclareMathAlphabet{\mathbcalboondox}{U}{BOONDOX-calo}{b}{n}

\newcommand{\no}{\nonumber}

\newcommand{\red}[1]{\textcolor{red}{#1}}

\newcommand{\cC}{\mathcal C}
\newcommand{\cE}{\mathcal E}

\newcommand{\cM}{\mathcal M}
\newcommand{\cO}{\mathcal O}

\newcommand{\cT}{\mathcal T}

\allowdisplaybreaks

\begin{document}

\title{Violation of Eigenstate Thermalization Hypothesis in Quantum Field Theories with Higher-Form Symmetry}
\author{Osamu Fukushima}
\email{osamu.f@gauge.scphys.kyoto-u.ac.jp}
\affiliation{Department of Physics, Kyoto University, Kyoto 606-8502, Japan}

\author{Ryusuke Hamazaki}
\email{ryusuke.hamazaki@riken.jp}
\affiliation{Nonequilibrium Quantum Statistical Mechanics RIKEN Hakubi Research Team, RIKEN Cluster for Pioneering Research (CPR), RIKEN iTHEMS, Wako, Saitama 351-0198, Japan}

\date{\today}

\begin{abstract}
We elucidate how the presence of  higher-form symmetries affects the dynamics of thermalization in isolated quantum systems.
Under reasonable assumptions, we analytically show that a $p$-form symmetry in a $(d+1)$-dimensional quantum field theory leads to the breakdown  of the eigenstate thermalization hypothesis (ETH) for many nontrivial $(d-p)$-dimensional observables. 
For discrete higher-form (i.e., $p\geq 1$) symmetry, this indicates the absence of thermalization for observables that are non-local but much smaller than the whole system size without any local conserved quantities.
We numerically demonstrate this argument for the (2+1)-dimensional $\mathbb{Z}_2$ lattice gauge theory.
While local observables such as the plaquette operator thermalize
even for mixed symmetry sectors,
the non-local observable exciting a magnetic dipole instead relaxes to the generalized Gibbs ensemble that takes account of the $\mathbb{Z}_2$ 1-form symmetry.

\end{abstract}

\maketitle

\paragraph{Introduction.--}
Symmetries play an essential role in 
statistical mechanics. When the system has local conserved quantities corresponding to symmetries, we should include them in the statistical ensemble. This fact has recently gained much attention in the thermalization of isolated quantum many-body systems~\cite{neumann1929beweis,deutsch1991quantum, srednicki1994chaos,PhysRevLett.80.1373,polkovnikov2011colloquium,eisert2015quantum, gogolin2016equilibration, d2016quantum, mori2018thermalization,ueda2020quantum}. While typical non-integrable systems without any symmetry locally relax to the canonical ensemble, existence of symmetries can affect its dynamics~\cite{PhysRevE.92.040103,hamazaki2016generalized,PhysRevResearch.2.033403,fratus2016eigenstate,PhysRevLett.130.140402,PhysRevE.107.014130,Borsi:2023nyt}. For example, integrable systems do not thermalize because of many symmetries but  relax to the generalized Gibbs ensemble (GGE)~\cite{rigol2007relaxation, rigol2009breakdown, iucci2009quantum, calabrese2011quantum, cassidy2011generalized, pozsgay2013generalized,pozsgay2014correlations,ilievski2015complete,  essler2016quench, vidmar2016generalized}, which takes account of  (quasi-)local conserved quantities.

The eigenstate thermalization hypothesis (ETH)~\cite{deutsch1991quantum, srednicki1994chaos, rigol2008thermalization} provides a sufficient condition for an isolated quantum system to thermalize and is extensively studied in various fields ranging from condensed matter~\cite{rigol2008thermalization,d2016quantum, mori2018thermalization} to holographic theory~\cite{Liska:2022vrd,deBoer:2016bov,Basu:2017kzo,Datta:2019jeo,Fitzpatrick:2015zha,Besken:2019bsu,Dymarsky:2019etq,Lashkari:2016vgj}. It states that every energy eigenstate becomes thermal for an observable of our interest and is verified in various non-integrable systems~\cite{rigol2008thermalization,santos2010onset,ikeda2011eigenstate,steinigeweg2013eigenstate, kim2014testing, beugeling2014finite, steinigeweg2014pushing, alba2015eigenstate, beugeling2015off, mondaini2017eigenstate, nation2018off, hamazaki2019random, khaymovich2019eigenstate,yoshizawa2018numerical, jansen2019eigenstate,PhysRevLett.126.120602,PhysRevLett.129.030602,neumann1929beweis, goldstein2010long, goldstein2010approach, goldstein2010normal, reimann2015generalization,hamazaki2018atypicality,sugimoto2023rigorous}. However, local conserved quantities  caused by, e.g., continuous global symmetry or integrability, break the ETH for the entire Hilbert space; in this case, the ETH is recovered after fixing conserved quantities~\cite{cassidy2011generalized, hamazaki2016generalized,PhysRevE.107.014130}.
In contrast, the influence of non-local conserved quantities caused by, e.g., discrete symmetries, is more subtle.
Indeed, they typically do not violate the (diagonal) ETH and thermalization of local observables~\cite{santos2010localization,hamazaki2016generalized,mondaini2016eigenstate} even when we consider mixed symmetry sectors, while they can sometimes be related to non-ergodicity~\cite{bernien2017probing, shiraishi2017systematic, turner2018weak, turner2018quantum, moudgalya2018exact, bull2019systematic, ho2019periodic, lin2019exact, schecter2019weak, shibata2020onsager,serbyn2021quantum,Sala:2019zru,PhysRevB.101.174204,PhysRevLett.124.207602,Feng:2021zuc,regnault2022quantum,PhysRevX.12.011050,PhysRevLett.129.090602,PhysRevB.106.214426}.

\begin{figure}[t]
\centering
\includegraphics[width=0.9\linewidth]{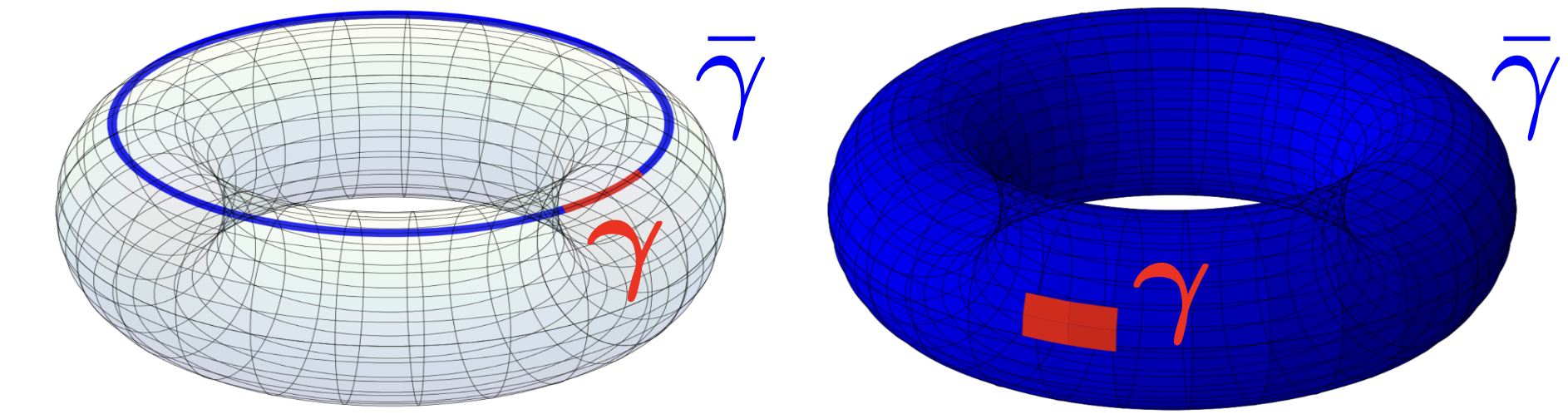}
\caption{Schematics of $p$-form symmetry and its influence on the eigenstate thermalization hypothesis (ETH).
The figure is for $\cM=T^2$ ($d=2$).
The blue and red regions respectively denote the areas where $U(\bar{\gamma})$ and $U(\gamma)$ nontrivially act, and the symmetry operator is given by $U(\tilde{C})=U(\bar{\gamma})U({\gamma})$.
If  $U(\gamma)$ satisfies the ETH with a  non-vanishing thermal average, the ETH for $U(\bar{\gamma})$ is broken.
(Left) The case with higher-form symmetry. The ETH-breaking operator $U(\bar{\gamma})$ is $(d-p)$-dimensional ($p=1$ is shown) and
 has a support much smaller than the size of the ``bath" $\cM\backslash\bar{\gamma}$.
 (Right) The case with the conventional symmetry ($p=0$), where the support of $U(\bar{\gamma})$ is comparable with the size of $\cM$.
} 
\label{fig:torus}
\end{figure}

Recently, {\it higher-form symmetry} has been proposed in the context of classifying phase structures of quantum field theories \cite{Gaiotto:2014kfa,Benini:2017dus,Gaiotto:2017yup,Tanizaki:2017bam,Komargodski:2017keh,Shimizu:2017asf,Gaiotto:2017tne,Kitano:2017jng,Gomis:2017ixy,Anber:2018iof,Hsin:2018vcg,Komargodski:2020mxz} (see Ref.~\cite{McGreevy:2022oyu} for applications to condensed matter) as a generalized concept of conventional global symmetries.
It is characterized by topological symmetry operators, whose correlation functions remain invariant under  continuous deformations.
The generalization is carried out concerning the dimensionality of charged objects and symmetry operators:
for $p$-form symmetries in a $(d+1)$-dimensional spacetime, the charged objects should be $p$-dimensional, and the symmetry operators are $(d-p)$-dimensional.
Conventional global symmetries  correspond to 0-form symmetries.

Despite the extensive research on static aspects of higher-form symmetry, its dynamical consequences are little understood.
Such higher-form symmetries nest intrinsically in gauge theories (e.g., Yang-Mills theories), which have been actively studied even in the condensed matter and atomic-molecular-optical contexts~\cite{Delacretaz:2019brr,Zhao:2020vdn,Pace:2023gye,Martinez:2016yna,Dai_2017,Kokail:2018eiw,Yang:2020yer,Klco:2018kyo,Gorg:2018xyc,Schweizer:2019lwx,Mil:2019pbt,Surace:2019dtp,deJong:2021wsd,Zhou:2021kdl,Wang:2022dpp,Tan:2019kya,zhang2017observation,Zohar:2021nyc}, not to mention high energy physics.
While a few studies exist to discuss thermalization dynamics in specific models with the generalized symmetries~\cite{Stahl:2023tkh,Stephen:2022kzd,khudorozhkov2022hilbert,Fagotti:2022npd,Buca:2023mri},  general consequences due to the higher-form symmetry have seldom been uncovered.


In this Letter,
we analytically show that  
nontrivial observables break the ETH 
when higher-form symmetries are present under certain assumptions (Fig.~\ref{fig:torus}).
In the case of discrete symmetry groups, the breakdown of the ETH is caused by non-local conserved quantities.
For a  $p$-form symmetry, such ETH-breaking operators become $(d-p)$-dimensional, which are non-local but have a  much smaller size than the entire system for $p\geq 1$. 
We demonstrate this statement for the two-dimensional $\mathbb{Z}_2$ lattice gauge theory with  $\mathbb{Z}_2$ 1-form symmetry~\cite{Gauss_foot}. 
Furthermore, while local observables relax to the canonical ensemble, the non-local 
operator exciting a magnetic dipole
instead relaxes to the GGE that considers the higher-form symmetry.
Our results indicate that symmetries cause nontrivial  thermalization processes revealed by non-local observables, which go beyond conventional statistical mechanics.



\paragraph{Higher-form symmetry.--}
We consider quantum field theories in a $(d+1)$-dimensional spacetime, $\cM\times\mathbb{R}$, where $\cM$ is a connected $d$-dimensional space manifold.
Let $G$ be an abelian group, and the system is supposed to have a $G$ $p$-form symmetry, i.e., there exists a $(d-p)$-dimensional topological operator $U_\alpha(\cC)$ $(\alpha\in G)$, where $\cC\subset \cM\times\mathbb{R}$ denotes a $(d-p)$-dimensional closed surface~\cite{topo_foot}.


We specifically consider the symmetry operator $U_\alpha(C)$ lying in the space for each fixed time, i.e., $C\subset \cM$. Then, the topological property of $U_\alpha(C)$ leads to $[U_\alpha(C),e^{-iH\delta t}]=0$ for an infinitesimal time slice $\delta t$, and thus $[H,U_{\alpha}(C)]=0$. 
Importantly, $U_\alpha(C)$ has a $(d-p)$-dimensional support $C$, which is non-local but much smaller than the entire $d$-dimensional system for $p\geq 1$. 
This contrasts with conventional 0-form symmetries, whose symmetry operator is $d$-dimensional, i.e., its support is comparable to the entire system (Fig.~\ref{fig:torus}).
While the following discussion holds both for continuous and discrete symmetries, we especially focus on discrete symmetry, which typically entails non-local conserved quantities alone.


Under this setting, the Hamiltonian is block-diagonalized by  $U_\alpha(C)$.
Then, the ETH for $U_\alpha(C)$  trivially  breaks down for the entire Hilbert space within the  energy shell, where symmetry sectors are mixed.
However, this does not necessarily indicate the breakdown of the ETH for other nontrivial observables since $U_\alpha(C)$ does not necessarily provide a local conserved quantity.
Indeed, there are several evidences~\cite{santos2010localization,hamazaki2016generalized,mondaini2016eigenstate} that the ETH holds for local observables despite the existence of the discrete  symmetry, especially the 0-form symmetry.
In that case,  eigenstate expectation values of those observables can be the same even for different symmetry sectors.
For example, the transverse-field Ising model $H_\mathrm{TFIM}=\sum_{\mathbf{R},\mathbf{R}'\in\cM}J_{\mathbf{R},\mathbf{R}'}\sigma^3(\mathbf{R})\sigma^3(\mathbf{R}')+\sum_{\mathbf{R}\in\cM}g_{\mathbf{R}}\sigma^1(\mathbf{R})$ has a $\mathbb{Z}_2$ 0-form symmetry $U(C=\cM)=\prod_{\mathbf{R}\in\cM}\sigma^1(\mathbf{R})$, where $\sigma^{1,2,3}(\mathbf{R})$ denote the Pauli matrices acting on the vertices $\mathbf{R}$. While we have two symmetry sectors with $U=\pm 1$, they will not lead to distinct eigenstate expectation values for typical local observables, say $\sigma^1(\mathbf{R})$, in the thermodynamic limit.


\paragraph{Breakdown of the ETH for nontrivial operators.--}
We now state our main result: higher-form symmetry of a non-degenerate Hamiltonian leads to the breakdown of the ETH even for many \textit{nontrivial} $(d-p)$-dimensional operators. For this purpose, we require the following reasonable assumptions:
%
i) the operator $U_\alpha(\tilde{C})$ can be decomposed as $U_\alpha(\tilde{C}) = U_\alpha(\gamma) U_\alpha(\bar{\gamma})$, where we have introduced a $(d-p)$-dimensional submanifold $\gamma\: (\subset \tilde{C})$ and its complement $\bar{\gamma}:= \tilde{C}\backslash \gamma$, both of which have boundaries.
ii) For at least one nontrivial closed surface, say $\tilde{C}\,(\subset \cM)$, 
the energy shell contains eigenstates in different symmetry sectors defined by $U_{\alpha}(\tilde{C})$.
iii) The microcanonical average $\braket{U_{\alpha}(\gamma)}_\mathrm{mc}^{\Delta E}(E)$ defined from the energy shell $[E,E+\Delta E]$ takes a nonzero value in the thermodynamic limit.

Under the above assumptions, we show that either $U_\alpha(\gamma)$ or $U_\alpha(\bar{\gamma})$ necessarily breaks the ETH~\cite{herm_foot} within the energy shell $[E,E+\Delta E]$ (see Supplemental Material~\cite{sup_foot} for a proof). 
Our result indicates that, while the discrete symmetry and topology may not affect the thermalization of conventional local observables, their effect significantly emerges in the dynamics of $(d-p)$-dimensional non-local objects.
We stress that, while such non-local observables go beyond conventional statistical mechanics, they have actively been studied since they play an essential physical role in gauge theory~\cite{Cordova:2022ruw,Gomes:2023ahz}. 
Furthermore, non-local operators  have  become accessible in state-of-the-art experiments using artificial quantum systems~\cite{endres2011observation,hilker2017revealing,semeghini2021probing}.


Let us point out the notable aspect of our results for higher-form symmetry with $p\geq 1$, although our results 
provide a hitherto unknown consequence even for the conventional symmetry $p=0$.
For $p=0$, the ETH-breaking operators (say, $U_\alpha(\bar{\gamma})$) are $d$-dimensional, and the volume of their support, $V_{\bar{\gamma}}$, is comparable with the volume of the ``bath" $V_{\cM\backslash\bar{\gamma}}$  for large system-size limit, i.e., $V_{\bar{\gamma}}/V_{\cM\backslash\bar{\gamma}}\rightarrow \mathrm{finite}$ (see Fig.~\ref{fig:torus}, right).
For the example of $H_\mathrm{TFIM}$, $\prod_{\mathbf{R}\in\cM\backslash\gamma}\sigma^1(\mathbf{R})$ breaks the ETH if $\prod_{\mathbf{R}\in\gamma}\sigma^1(\mathbf{R})$ satisfies it.
Thus, the breakdown of the ETH might also be attributed to the smallness of the bath.
In contrast, for higher-form symmetry with $p\geq 1$, we have $V_{\bar{\gamma}}/V_{\cM\backslash\bar{\gamma}}\rightarrow 0$ in the thermodynamic limit (Fig.~\ref{fig:torus}, left).
Thus, the higher-form symmetry hinders thermalization even when the bath is regarded as much larger than the support of the observable of our interest.

Our main claim is generalized to an operator $ U(\bar{\gamma}) A(g)^\dagger$, where $A(g)$ is an operator defined on an arbitrary region $g\:(\subset\cM)$ satisfying $g\cap \bar{\gamma}=\phi$.
That is,
$A(g) U(\gamma)$ or $ A(g)^\dagger U(\bar{\gamma}) $ violates the ETH if we impose an assumption iii)' $\langle A(g) U(\gamma) \rangle_{\rm mc}^{\Delta E}(E)\neq 0 $ instead of iii).
This generalization indicates that for a fixed $\gamma$, we  have many ETH-violating operators corresponding to the choice of $g$ and $A(g)$.
Note that, while $g\subseteq\gamma$ for 0-form symmetries, $g$ may not be included in (or have even larger dimension than) $\gamma$ for higher-form symmetries.

Finally,  the symmetry can, in turn, \textit{ensure} the ETH~\cite{eth_foot} for certain operators.
Indeed, the so-called charged operators $W$,
for which $U_\alpha(C)WU_\alpha^{-1}(C)=e^{i\alpha w}W$ holds with some charge $w$,
satisfy $\braket{E_n|W|E_n}=0$ for all $n$ when $e^{i \alpha w}\neq 1$.
Then, $W$ satisfies the ETH, and the long-time average of $\braket{W(t)}$ becomes zero.
For $H_\mathrm{TFIM}$ with 0-form symmetry, $W=\sigma^3(\mathbf{R})$ satisfies this condition. For the higher-form symmetry, the Wilson line in the lattice gauge theory discussed later satisfies this condition.



\paragraph{$\mathbb{Z}_2$ lattice gauge theory.--}

\begin{figure}[t]
\centering
\includegraphics[width=1.\linewidth]{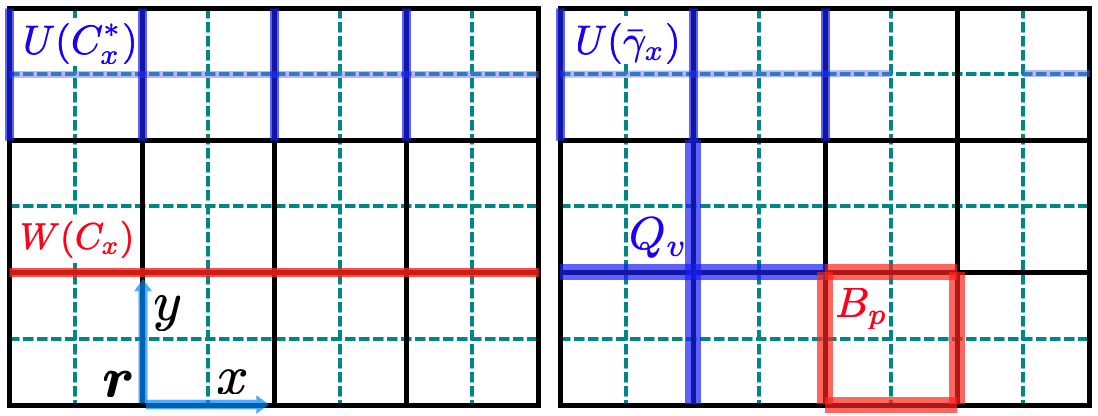}
\caption{Schematic diagram of our lattice model ($N_x=4$ and $N_y=3$ are shown) under the  periodic boundary condition.
The solid and green lines denote the lattice and the dual lattice, respectively.
We illustrate examples of the operators $Q_v, U(C_x^*),W(C_x), U(\bar{\gamma}_x)$, and $B_p$, where  
the Pauli matrices $\sigma^3$ and $\sigma^1$ respectively act on the red and blue links.
}\label{fig:lattice}
\end{figure}

We demonstrate the general discussion above using the
(2+1)-dimensional $\mathbb{Z}_2$ lattice gauge theory on a square lattice forming a 2-torus $\cM=T^2$.
The Hamiltonian $H_{\mathbb{Z}_2}$ is given by \cite{Kogut:1974ag,Fradkin:1978th,Fradkin:1978dv}
\begin{align}
- 
 \sum_{\bm{r}}J_{\bm{r},xy} 
\sigma_{\bm{r},x}^3\sigma_{\bm{r}+\bm{e}_x,y}^3
\sigma_{\bm{r}+\bm{e}_y,x}^3\sigma_{\bm{r},y}^3
-\sum_{\bm{r},j}\sigma_{\bm{r},j}^1,
\label{Z2-Hamiltonian}
\end{align}
where $\sigma_{\bm{r},j}^{1,2,3}$ denote the Pauli matrices acting on the link $(\bm{r},j)$,
which is specified by the coordinate of vertices $\bm{r}$ and the direction $j=x,y$.
This system has a $\mathbb{Z}_2$ 1-form symmetry, and the spatial symmetry operators are characterized by $H_1(T^2,\mathbb{Z}_2)=\mathbb{Z}_2\oplus\mathbb{Z}_2$~\cite{Roumpedakis:2022aik}.
Indeed, the system has two independent symmetry operators corresponding to the $x$-cycle and $y$-cycle.


To remove the residual gauge redundancies after the temporal gauge-fixing\;\cite{Fradkin:1978th}, we project the entire Hilbert space onto the physical one.
Here, spatial gauge transformation is generated by the local operator
$
Q_{v}:=\prod_{\substack{b:\mbox{ \scriptsize spatial link}, b\ni v}}\sigma_{b}^1,$ where $v$ denotes the vertex.
The operator $Q_v$  satisfies $Q_v^2=1$,  $[H_{\mathbb{Z}_2},Q_v]=0$.
Then, the physical Hilbert space is given by~\cite{Fradkin:1978th}
$
\operatorname{span}\left\{ |\psi\rangle\;\big|\; Q_v|\psi\rangle = +|\psi\rangle ,\; {}^{\forall} v:\mbox{ vertices} \right\},
$
where the constraint can be regarded as the $\mathbb{Z}_2$ analog of the Gauss law. 
After this projection, the expectation value of a non-gauge invariant operator for physical states $|\psi\rangle$ always vanishes.


We next define the 't Hooft and Wilson operators on the spatial directions as~\cite{tHooft:1977nqb,Ukawa:1979yv,Shimazaki:2020byo}
\begin{align}
U(C^*)=&\, \prod_{b^*\in C^*}\sigma_{b^{*}}^1 = U^{-1}(C^*),
\end{align}
and $W(C)=\prod_{b\in C}\sigma_{b}^3 = W^{-1}(C)$.
Here, $C$ and $C^*$ are closed loops on the lattice and dual lattice, respectively (see Fig.~\ref{fig:lattice}).
Both $U(C^*)$ and $W(C)$ commute with the operator $Q_v$ and thus are gauge invariant.

\begin{figure*}[t]
\centering
\includegraphics[width=1.0\linewidth]{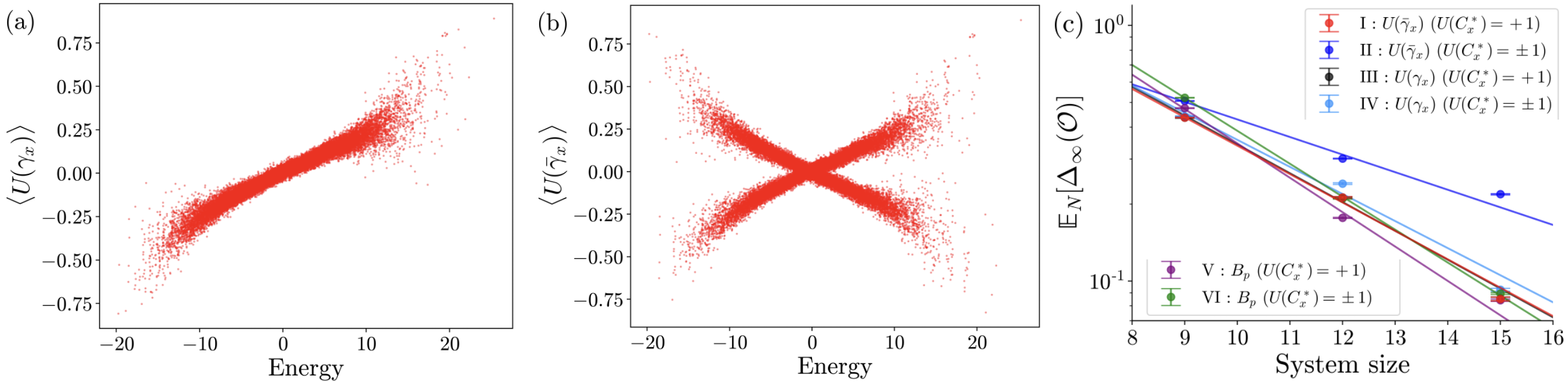}
\caption{ Expectation values for energy eigenstates in the $5\times 3$ lattice.
(a) The local observable $U(\gamma_x)$ satisfies the ETH.
(b) The observable with a one-dimensional support $U(\bar{\gamma}_x)$ violates the ETH.
The expectation values are separated into two sectors classified by the value of 1-form symmetry, i.e., $U(C_x^*)=\pm 1$.
(c) System-size dependence of the ETH measure $\Delta_{\infty}$.
The decay for $U(\bar{\gamma}_x)$ with the total symmetry sectors (Case II) is much slower than the other cases, which indicates that the ETH is hindered.
The fitting parameters are shown in Supplemental Material~\cite{sup_foot}.
}
\label{fig:ETH}
\end{figure*}

The 't Hooft operator $U(C^*)$ serves as the $\mathbb{Z}_2$ 1-form symmetry operator of this model, satisfying
$
[H_{\mathbb{Z}_2}, U(C^*) ] =0.
$
This operator is topological since it satisfies
$
U(C_1^*)|\psi\rangle = U(C_2^*)|\psi\rangle
\label{U-spatial}
$
if $C_1^*$ and $C_2^*$ are homotopically equivalent.
It follows that $U(C^*)|\psi\rangle=|\psi\rangle$ if the dual closed loop $C^*$ is topologically trivial, i.e., it can be continuously deformed to a point.


The 't Hooft operator $U(C^*)$  measures the ``electric'' charge of the Wilson operator.
We  define closed loops on the lattice winding around the $x$-/$y$-cycle by $C_x$ and $C_y$ (and similarly the loops on the dual lattice by $C_x^*$ and $C_y^*$).
Then, the operators $W$ and $U$ satisfy 
$
U(C_i^*)W(C_j)U^{-1}(C_i^*) = (-1)^{\delta_{ij} + 1}W(C_j),
$
which is  operator-realization of the electric $\mathbb{Z}_2$ 1-form symmetry~\cite{Pace:2023gye}.


\paragraph{ETH breaking by $\mathbb{Z}_2$ 1-form symmetry.--}
Let us  demonstrate the  violation of the ETH for the $\mathbb{Z}_2$ lattice gauge theory.
We take the coupling constants $J_{\mathbf{r},xy}$ in (\ref{Z2-Hamiltonian})  to be weakly random (i.e., $J_{\bm{r},xy}$ is uniformly chosen from $[0.7, 0.8]$) to avoid unwanted degeneracy and integrability.
We consider the $N_x\times N_y$ 2-torus and define $x/y$-cycles for the lattice and dual lattice as $C_{x/y}$ and $C^*_{x/y}$, respectively (Fig.~\ref{fig:lattice}).

We calculate the eigenstate  expectation values of local operators $U(\gamma_x)$, $B_p$,  and nonlocal one-dimensional observable $U(\bar{\gamma}_x)$.
Here, $\gamma_x$ is just one link included in $C^*_x$, and $\bar{\gamma}_x:=C_x^*\backslash \gamma_x$. 
The operators $U(\gamma_{x})$ and $U(\bar{\gamma}_{x})$ represent a magnetic dipole excitation residing at the endpoints of $\gamma_x$~\cite{Fradkin:1978th}.
The plaquette operator $B_p$ is defined by
$
B_p:= \prod_{b\in \mbox{\scriptsize plaquette }p}\sigma_{b}^3.
 $
Figure~\ref{fig:ETH}(a) shows  that the local observable $U(\gamma_x)$ satisfies the ETH.
In contrast,
Fig.~\ref{fig:ETH}(b)
demonstrates that the non-local one-dimensional observable $U(\bar{\gamma}_x)$ has two branches of the eigenstate-expectation values,
indicating the breakdown of the ETH
owing to the general mechanism explained above.
The ETH is recovered when we consider eigenstates within the symmetry sector for $U(C_x^*)=1\:(-1)$, even without separating the sector for $U(C_y^*)$.
The result for $B_p$ is given in the Supplemental Material~\cite{sup_foot}.

To test the ETH more quantitatively, we perform the finite-size scaling analysis.
We define the deviation measure for an observable $\cO$ \cite{PhysRevLett.126.120602} by
$
 \Delta_{\infty}(\cO) := 
 \max_{\substack{ n , E_n\in [E,E + \delta E]}} 
 \left| \langle E_n | \cO | E_n \rangle - \langle \cO \rangle_{\rm mc}^{\delta E}(E_n) \right|,
$
where $E_n=\langle E_n |H_{\mathbb{Z}_2}| E_n\rangle$ is an energy eigenvalue.
The strong ETH corresponds to  $\Delta_\infty\rightarrow 0$ in the thermodynamic limit.
Furthermore, $\Delta_\infty$ is expected to decay exponentially $\sim e^{-s(E)N/2}$ for a fully chaotic system, where $s(E)$ is the entropy density at energy $E$~\cite{d2016quantum,PhysRevLett.126.120602}. 
Figure 3(c) shows the system-size dependence of the disorder-averaged measure $\mathbb{E}[\Delta_\infty(\mathcal{O})]$, which is fitted with a function $e^{-aN+b}$.
First, the local observables $B_p$ and $U(\gamma_x)$ (irrespective of whether we resolve the symmetry sector) and the non-local observable $U(\bar{\gamma}_x)$ after resolving the symmetry sector exhibit sufficiently fast exponential decay with a relatively similar rate.
This indicates the ETH for these observables. 
In contrast, $U(\bar{\gamma}_x)$ for the total symmetry sector decays much slower than the other cases, though it keeps decreasing due to the finite-size effect.
Combining the general argument and the ETH for $U(\gamma_x)$, we conclude that the ETH for $U(\bar{\gamma}_x)$ breaks down due to the higher-form symmetry.

Note that the Wilson line always satisfies
$
\braket{E_n|W(C_{x/y})|E_n}=0
$
for all $n$, because of the general discussion for the charged operator discussed previously.
Consequently, the long-time average of the Wilson operator always vanishes.

\paragraph{GGE with $\mathbb{Z}_2$ 1-form symmetry.--}
We next argue that the stationary value of an observable that is non-local in the $x$-direction but local in the $y$-direction (e.g., $U(\bar{\gamma}_x)$) is described by the GGE
that takes account of the $\mathbb{Z}_2$ 1-form symmetry.
That is, we have $\langle \cO \rangle_{\rm GGE} = 
\mathrm{Tr}[\cO \, \rho_\mathrm{GGE}(\beta,\lambda_x,\mu_x )]$ with
\begin{align}
\rho_\mathrm{GGE}=\frac{1}{Z_\mathrm{GGE}}e^ {-\beta H_{\mathbb{Z}_2} 
-   \lambda_x U(C^*_x) - \mu_x U(C^*_x) H_{\mathbb{Z}_2}  },
\label{GGE-definition}
\end{align}
where ``chemical potentials" $\lambda_x$ and $\mu_x$ are uniquely determined from the initial values of the conserved quantities $H_{\mathbb{Z}_2},U(C_x^*),$ and $U(C_x^*)H_{\mathbb{Z}_2}$, and $Z_\mathrm{GGE}$ is the normalization constant. 
Our GGE is justified as the stationary state if we assume the restricted ETH for each $U(C_x^*)$-symmetry sector for the observable $\cO$ (See Supplemental Material~\cite{sup_foot}).


Figure~\ref{fig:time-evolution} shows time evolutions of $\mathcal{O}=U({\gamma}_x)$ and $U(\bar{\gamma}_x)$.
For $U({\gamma}_x)$, the stationary value is well described by the canonical ensemble $\rho_\mathrm{can}=Z_\mathrm{can}^{-1}e^{-\beta_\mathrm{can}H_{\mathbb{Z}_2}}$, where $\mathrm{Tr}[{U(\gamma_x)\rho(t)}]\simeq \mathrm{Tr}[{U(\gamma_x)\rho_\mathrm{can}}]$
holds for most of the time.
In contrast, the canonical ensemble fails for $U(\bar{\gamma}_x)$.
Instead, 
$\mathrm{Tr}[U(\bar{\gamma}_x)\rho(t)]\simeq\mathrm{Tr}[U(\bar{\gamma}_x)\rho_\mathrm{GGE}]$ holds for most of the time.
We stress that $\rho_\mathrm{GGE}$ works well even though we do not consider the effect of $U(C_y^*)$, probably because $U(\bar{\gamma}_x)$ is local in the $y$-direction.


Note that the GGE suitable for a general finite abelian group $G$ is obtained by assuming the restricted ETH for each symmetry sector as 
$
\rho_\mathrm{GGE}^G=e^{-\beta^G H - \sum_j \lambda_j^G P_j - \sum_j \mu_j^G P_j H},
\label{GGE-general}
$
where $P_i$ are the projections to each symmetry sector.
For symmetry sectors defined by $U(C_x^*)$ with $G=\mathbb{Z}_2$, this ensemble is indeed equivalent to (\ref{GGE-definition}) after redefinitions of the chemical potentials.

\begin{figure}[t]
\centering
\includegraphics[width=1.0\linewidth]{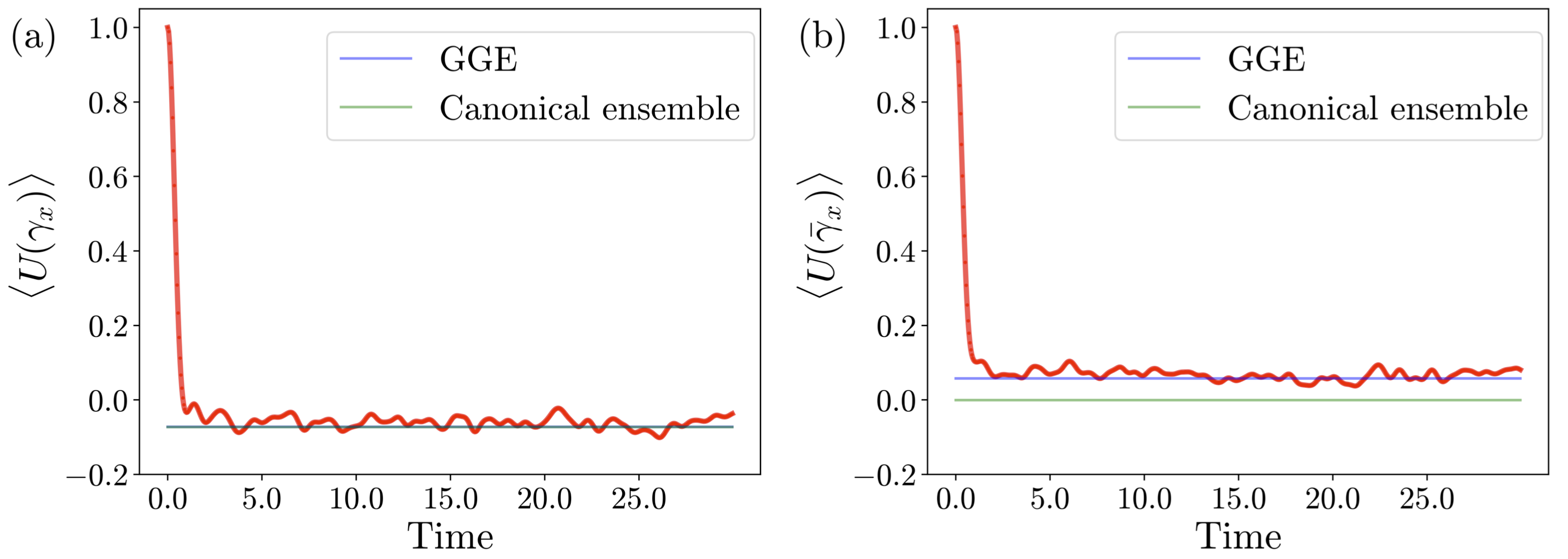}
\caption{Time evolution of the expectation values of $U(\gamma_x)$ and $U(\bar{\gamma}_x)$
for the $4\times 3$ lattice.
The blue and  green lines indicate the prediction of the GGE in Eq.~\eqref{GGE-definition} and the standard canonical ensemble, respectively.
For the local observable $U(\gamma_x)$ (left), the stationary state is described by the canonical ensemble, which is almost overlapping with the GGE result.
In contrast, the stationary value of $U(\bar{\gamma}_x)$ (right) differs from the canonical ensemble and is described by the GGE.
The initial states are random superpositions of eigenstates of $U(\gamma_x)$ (left) or $U(\bar{\gamma}_x)$ (right) with the eigenvalue $+1$, whose energy expectations lie within  $[-5.0,-3.0]$.
}\label{fig:time-evolution}
\end{figure}


\paragraph{Conclusion and outlook.--}
We analytically show that the existence of a $p$-form symmetry leads to the ETH-violation of many $(d-p)$-dimensional observables in the form of  $U_\alpha(\bar{\gamma})$ under certain assumptions.
 A significant feature of this statement is that the ETH-violating observable has a non-local but  lower-dimensional support rather than the whole $d$-dimensional space manifold for $p\geq 1$.
This implies that such  objects can be described by the suitable GGE instead of the canonical ensemble.
We use the $\mathbb{Z}_2$  lattice gauge theory to demonstrate the above statements.
The discussion on the breakdown of the ETH can be applied to 
systems with $p$-form symmetries, e.g., various quantum field theories such as $SU(N)$ Yang-Mills theory with center symmetries.

Our results indicate that symmetries cause nontrivial  thermalization dynamics for non-local observables, which go beyond conventional statistical mechanics.
We stress that this ETH-violation stably holds even under local perturbations to the Hamiltonians because the higher-form symmetry is robust against them.
For future direction, higher-form symmetry may impact entanglement structures of certain subsystems, including entanglement-entropy dynamics, which is a non-local quantity. As entanglement entropy is vital in holography, exploring it may illuminate black-hole dynamics through gauge/gravity duality.

\begin{acknowledgments}
We thank Y. Hidaka for fruitful discussions.
We also thank T. Fukuhara for informing us of some references on experimental observations of non-local observables. We are also grateful to  MACS program in Kyoto University for stimulating our collaborations.
O.F. was supported by Grant-in-Aid for JSPS Fellows No.~21J22806.
R.H. was supported by JST ERATO-FS Grant Number JPMJER2204, Japan.
\end{acknowledgments}

\bibliographystyle{apsrev4-1}
\bibliography{higher_form_ETH}

\begin{thebibliography}{141}
\bibitem [{top()}]{topo_foot}%
  \BibitemOpen
  \href@noop {} {}\bibinfo {note} {Here, an operator $U_\alpha(\cC)$ is said to
  be topological if and only if the transition amplitude satisfies $\langle f |
  \cT ( U_\alpha(\cC) e^{-i\int_{t_i}^{t_f}Hdt}) | i \rangle = \langle f | \cT
  ( U_\alpha(\cC') e^{-i\int_{t_i}^{t_f}Hdt}) | i \rangle$ for arbitrary
  initial and final states $|i\rangle$ and $|f\rangle$, and homotopically
  equivalent closed surfaces $\cC' \subset \cM\times \mathbb{R}$, where $\cT$
  denotes the time-ordering.}\BibitemShut {Stop}%
\bibitem [{her()}]{herm_foot}%
  \BibitemOpen
  \href@noop {} {}\bibinfo {note} {Note that $U_\alpha(\gamma)$ and
  $U_\alpha(\bar{\gamma})$ are in general non-Hermitian. However, the breakdown
  of the ETH for these operators leads to that of certain Hermitian operators
  as well. Indeed, if $U_\alpha(\bar{\gamma})$ breaks the ETH, we can show that
  either of the Hermitian operators
  $U_\alpha(\bar{\gamma})+U_\alpha^\dag(\bar{\gamma})$ or
  $i(U_\alpha(\bar{\gamma})-U_\alpha^\dag(\bar{\gamma}))$ breaks the
  ETH.}\BibitemShut {Stop}%
\bibitem [{eth()}]{eth_foot}%
  \BibitemOpen
  \href@noop {} {}\bibinfo {note} {In this manuscript, we only consider the
  diagonal ETH and do not discuss the off-diagonal
  ETH~\cite{PhysRevLett.111.050403}, which states that the off-diagonal matrix
  elements with respect to the energy eigenstates vanish in the thermodynamics
  limit.}\BibitemShut {Stop}%
\end{thebibliography}

\end{document}


\maketitle


\section{Proof of the breakdown of the ETH under symmetries}

As discussed in the main text,  higher-form symmetry of a non-degenerate Hamiltonian leads to the breakdown of the eigenstate thermalization hypothesis (ETH) even for many {nontrivial} operators. 
We here show this fact.
The case $A(g)=1$ in the following corresponds to the main claim in the main text, and the ETH-violating operators are $(d-p)$-dimensional in this case.

As stated in the main text, we require the following reasonable assumptions:
i) the topological operator $U_\alpha(\tilde{C})$ can have boundaries, i.e., $U_\alpha(\gamma)$ with an arbitrary $(d-p)$-dimensional submanifold $\gamma\: (\subset \tilde{C})$ is a well-defined (not-null) operator.
One can then decompose the operator as $U_\alpha(\tilde{C}) = U_\alpha(\gamma) U_\alpha(\bar{\gamma})$, where we have introduced the complement of $\gamma$ as  $\bar{\gamma}:= \tilde{C}\backslash \gamma$.
ii) For at least one nontrivial closed surface, say $\tilde{C}\,(\subset \cM)$, there are energy eigenstates $|E_n\rangle$, $|E_m\rangle$ with $E_n,E_m\in[E,E+\Delta E]$ such that $\langle E_n | U_{\alpha}(\tilde{C}) | E_n \rangle\neq \langle E_m | U_{\alpha}(\tilde{C}) | E_m \rangle$.
In other words, the energy shell of our interest contains eigenstates in different symmetry sectors defined by $U_{\alpha}(\tilde{C})$.
iii)' The microcanonical average $\braket{A(g) U_{\alpha}(\gamma)}_\mathrm{mc}^{\Delta E}(E)$ of the operator $A(g) U_\alpha(\gamma)$ defined from the energy shell $[E,E+\Delta E]$ takes a nonzero value in the thermodynamic limit.
The operator $A(g)$ here is defined on an arbitrary region $g\:(\subset \cM)$ that satisfies $g\cup \bar{\gamma}=\phi$ (empty region).
If $A(g)$ is the identity operator, the assumption iii)' just reduces to iii).

With these assumptions, we can show that either $A(g) U_\alpha(\gamma)$ or $U_\alpha(\bar{\gamma}) A(g)^\dagger$ necessarily breaks the ETH within the energy shell $[E,E+\Delta E]$. 
To see this, we consider a $(d-p)$-dimensional surface $\gamma$ with boundary, which satisfies the property iii)'.
If the operator $A(g) U_{\alpha}(\gamma)$ does not satisfy the ETH, our claim holds; we thus consider the case where $A(g) U_{\alpha}(\gamma)$ satisfies the ETH, i.e., 
\begin{align}
\langle E_n | A(g) U_{\alpha}(\gamma) | E_n \rangle\simeq \langle E_m | A(g) U_{\alpha}(\gamma) | E_m \rangle\simeq \braket{A(g) U_{\alpha}(\gamma)}_\mathrm{mc}^{\Delta E}(E).
\end{align}
Since $H$ is assumed to have no degeneracy, its eigenstates $|E_n\rangle$, $|E_m\rangle$ are also eigenstates of $U_{\alpha}(\tilde{C})$.
Since the group $G$ is abelian,  the eigenvalues are expressed as
\begin{align}
U_{\alpha}(\tilde{C})|E_n\rangle=e^{i\alpha q_n}|E_n\rangle
\end{align}
 and 
 \begin{align}
U_{\alpha}(\tilde{C})|E_m\rangle=e^{i\alpha q_m}|E_m\rangle,
\end{align}
where $q_n,q_m\in\mathbb{R}$. From the assumption ii), we can assume that $|E_n\rangle$ and $|E_m\rangle$ belong to different sectors, i.e., $e^{i\alpha q_n}\neq e^{i\alpha q_m}$.

Now, the definition of $\bar{\gamma}$ indicates 
\begin{align}
\langle E_n |A(g) U_{\alpha}^{-1}(\bar{\gamma}) | E_n\rangle = \langle E_n |A(g) U_{\alpha}(\gamma)U_{\alpha}(\tilde{C})^{-1} | E_n\rangle = e^{-i\alpha q_n}\langle E_n | A(g) U_{\alpha}(\gamma)| E_n\rangle
\end{align}
and 
$\langle E_m |A(g) U_{\alpha}^{-1}(\bar{\gamma}) | E_m\rangle = e^{-i\alpha q_m}\langle E_m |A(g) U_{\alpha}(\gamma)| E_m\rangle$.
Recalling the assumption of the ETH and iii)', i.e., 
\begin{align}
\langle E_n | A(g) U_{\alpha}(\gamma) | E_n \rangle\simeq \langle E_m | A(g) U_{\alpha}(\gamma) | E_m \rangle\simeq \braket{A(g) U_{\alpha}(\gamma)}_\mathrm{mc}^{\Delta E}(E)\neq 0,
\end{align}
we obtain the relation
$\langle E_n |A(g) U_{\alpha}^{-1}(\bar{\gamma}) | E_n\rangle \neq \langle E_m | A(g) U_{\alpha}^{-1}(\bar{\gamma}) | E_m\rangle.$ Finally, taking the complex conjugate, we have
\begin{align}
\langle E_n | U_{\alpha}(\bar{\gamma}) A(g)^\dagger | E_n\rangle \neq \langle E_m | U_{\alpha}(\bar{\gamma}) A(g)^\dagger | E_m\rangle.
\label{ETH-violation}
\end{align}
Thus, $U_{\alpha}(\bar{\gamma}) A(g)^\dagger$ breaks the ETH, and our claim has been proven.

\section{Numerical details of the ETH for the $\mathbb{Z}_2$ gauge theory}

\subsection{Fitting parameters for the finite-size scaling analysis}

Figure 3(c) in the main text shows the finite-size scaling of the deviation measure 
\begin{align}
\Delta_\infty(\cO) :=
\max_{\substack{ n , E_n\in [E,E + \delta E]}} 
 \left| \langle E_n | \cO | E_n \rangle - \langle \cO \rangle_{\rm mc}^{\delta E}(E_n) \right|.
\end{align}
The measure $\mathbb{E}[\Delta_\infty(\mathcal{O})]$ averaged over disorder is fitted with a function $e^{-aN+b}$.
Table S-1 shows the fitting parameters for each case. The decay for $U(\bar{\gamma}_x)$ with the total symmetry sectors (Case II) is much slower than the other cases, which indicates that the ETH is hindered only for this case.
We use the square lattices $(N_x,N_y)=(3,3),\,(4,3),\,(5,3)$ with $10^4$, $10^3$, and $10^2$ samples, respectively.
The energy window is set as $[E,E+\delta E]=[-1.3\, N_x,-1.0\, N_x]$.

\begin{table}[htbp]
\centering
\caption{Observables, symmetry sectors, and the fitting parameters $a$ and $b$ for six different cases.}
\begin{tabular}{ccccc}\hline
    Case & Observable & Symmetry sector & $a$ & $b$  \\ \hline\hline
    I & \multirow{2}{*}{$U(\bar{\gamma}_x)$} & $U(C^*_x)=1$ & $0.256\pm 0.017$ & $1.47 \pm 0.16$ \\
    II &  & $U(C^*_x)=\pm1$ & $0.158\pm 0.020$ & $0.74 \pm 0.20$ \\ \hline
    III & \multirow{2}{*}{$U({\gamma}_x)$} & $U(C^*_x)=1$ & $0.260\pm 0.021$ & $1.52 \pm 0.23$ \\
    IV &  & $U(C^*_x)=\pm1$ & $0.244\pm 0.036$ & $1.41 \pm 0.40$ \\ \hline
    V & \multirow{2}{*}{$B_p$} & $U(C^*_x)=1$ & $0.310\pm 0.024$ & $2.03 \pm 0.24$ \\
    VI &  & $U(C^*_x)=\pm1$ & $0.297\pm 0.005$ & $2.02 \pm 0.05$ \\
    \hline
\end{tabular}
\end{table}

\subsection{ETH for the plaquette operator}
Here, we show the detail of the ETH for a plaquette operator $B_p$ along the same line as those for $U(\gamma_x)$ and $U(\bar{\gamma}_x)$
in the main text.
Though this operator can be regarded as the minimal Wilson operator, it acts nontrivially on the links around one plaquette $p$. In this sense, the operator $B_p$ can be regarded as a local observable.

\begin{figure}[htbp]
\centering
\begin{minipage}[ht]{0.49\linewidth}
\centering
\includegraphics[width=1.\linewidth]{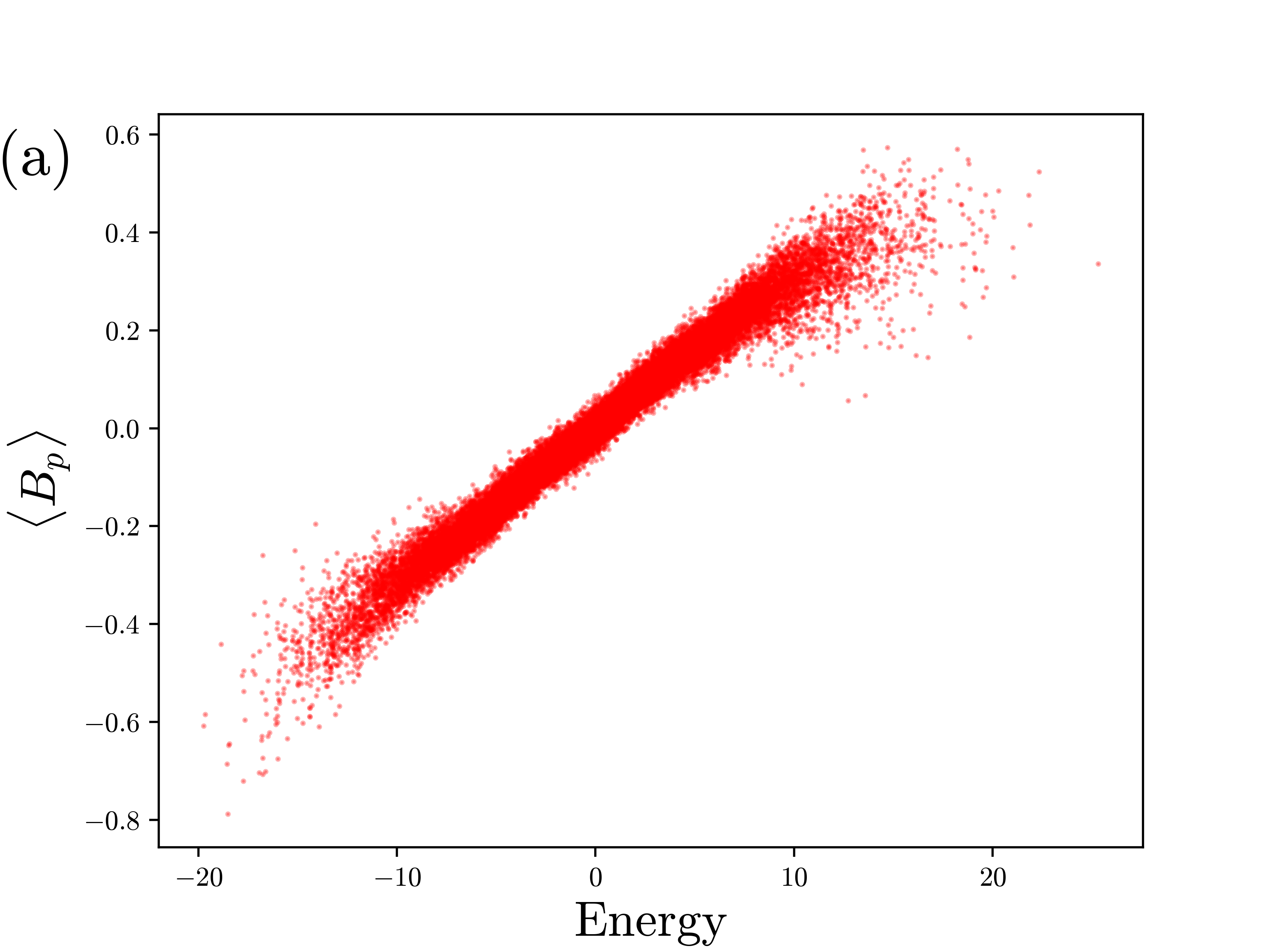}
\end{minipage}
\begin{minipage}[ht]{0.49\linewidth}
\centering
\includegraphics[width=1.\linewidth]{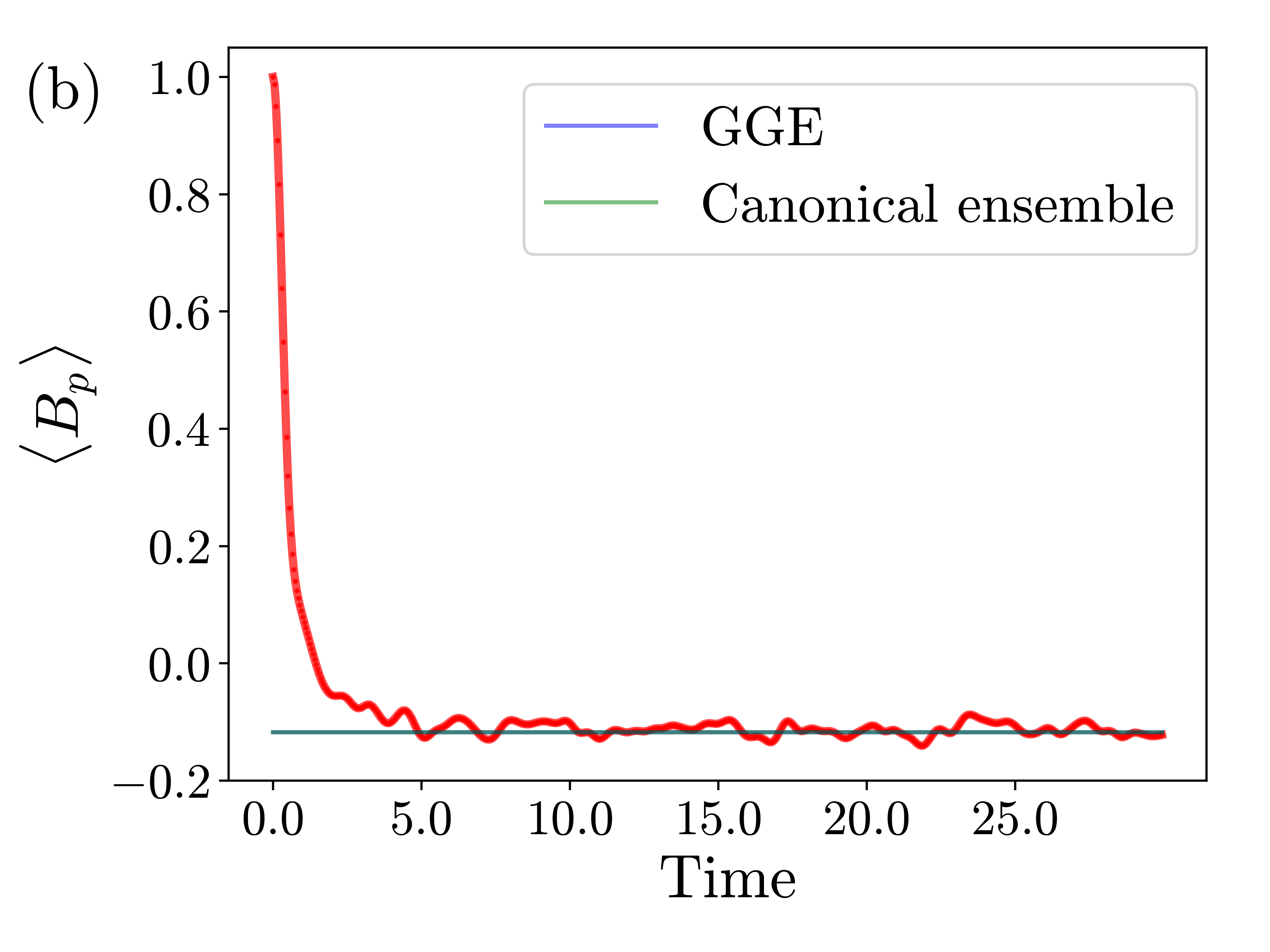}
\end{minipage}
\caption{(a) Expectation values of the local operator $B_p$ with respect to energy eigenstates for the $5\times 3$ lattice.
We see that  $B_p$ satisfies the ETH (also see Fig.~3(c) in the main text), 
even without resolving the symmetry sectors associated with $U(C^*_x)$ and $U(C^*_y)$.
(b) Time evolution of the expectation value of $B_p$ for the $4\times 3$ lattice.
The initial state is a random superposition of the eigenstates of $B_p$ with the eigenvalue $+1$, whose energy expectations lie within an energy window $E\in [-5.0,-3.0]$.
The prediction for  the GGE and that for canonical ensemble make no difference (i.e., two results are almost overlapped), and the stationary value of  $\langle B_p\rangle$ is described by them.
}
\label{fig:Bp}
\end{figure}

The expectation values versus eigenenergies are shown in Fig. \ref{fig:Bp}(a).
Along with Fig.~3(c) in the main text, we conclude that the operator $B_p$ satisfies the ETH even without resolving symmetry sectors.
Figure~\ref{fig:Bp}(b) shows the time evolution of the expectation value $\langle B_p(t)\rangle$.
The stationary state is well described by the ensemble average of the generalized Gibbs ensemble (GGE) and that of the canonical ensemble, where both ensembles provide almost the same prediction.

\section{Justification of the GGE}
\subsection{Justification for the case with $H_{\mathbb{Z}_2}$}
In the main text, we have introduced the GGE that takes into account discrete 1-form symmetry, $U(C_x^*)$.
Here we show that this GGE can be justified if we assume the ETH restricted for each symmetry sector with respect to $U(C_x^*)$.
Our GGE that takes account of the $\mathbb{Z}_2$ 1-form symmetry is given by 
\begin{align}\begin{split} 
\langle \cO \rangle_{\rm GGE} =&\, \operatorname{Tr} (\cO \rho_{\rm GGE}) ,\\
\rho_\mathrm{GGE}(\beta,\lambda_x,\mu_x )
:=&\, \frac{1}{Z_{\rm GGE}(\beta,\lambda_x,\mu_x )}
e^ {-\beta H_{\mathbb{Z}_2} 
-    \lambda_x U(C^*_x) - \mu_x U(C^*_x) H_{\mathbb{Z}_2}  },\\
Z_{\rm GGE}(\beta,\lambda_x,\mu_x ) :=&\,
\operatorname{Tr}[ e^ {-\beta H_{\mathbb{Z}_2} 
-    \lambda_x U(C^*_x) -\mu_x U(C^*_x) H_{\mathbb{Z}_2}  }],
\end{split}
\label{sup:GGE-definition}
\end{align}
where $\beta,\lambda_x,\mu_x$ are determined from the initial values of the conserved quantities $H_{\mathbb{Z}_2}, U(C_x^*)$, and $U(C_x^*)H_{\mathbb{Z}_2}$ (see Eq.~\eqref{GGEcondition}).

To see that the GGE describes the stationary state, we first obtain the expression for the stationary state, i.e., the long-time average of the expectation value of an observable $\cO$ (we assume that the temporal fluctuation around the long-time average is negligible in the thermodynamic limit).
For this purpose, we define the projection operator onto each symmetry sector by
\begin{align}
    P_{\pm}^{x} := \frac{1}{2}\big( 1 \pm U(C_{x}^*) \big),
    \label{projection}
\end{align}
Once we assume the ETH with respect to each symmetry sector, 
the expectation values $\langle P_{+}^{x} \rangle$, $\langle P_{+}^{x} H_{\mathbb{Z}_2} P_{+}^{x} \rangle$ and $\langle P_{-}^{x} H_{\mathbb{Z}_2} P_{-}^{x} \rangle$ suffice to specify the stationary state.
To see this, we consider the time evolution from  an initial state $\ket{\psi}=\sum_{n}c_n \ket{E_n}$ with the energy eigenstates $\ket{E_n}$.
Then, if we assume no degeneracy for the energy eigenvalues, the long-time average $\overline{\braket{\psi|\cO|\psi}}:=\lim_{T\rightarrow\infty}\frac{1}{T}\int_0^Tdt \braket{\psi(t)|\cO|\psi(t)}$ is given by 
\begin{align}
    \overline{\braket{\psi|\cO|\psi}} =&\,
    \sum_{n}|c_n|^2 \braket{E_n|\cO|E_n}\no\\
    =&\, \sum_{n_+:U(C_x^*)=+1\text{ sector}}|c_{n_+}|^2 \braket{E^+_{n_+}|\cO|E^+_{n_+}} + \sum_{n_-:U(C_x^*)=-1\text{ sector}}|c_{n_-}|^2 \braket{E^-_{n_-}|\cO|E^-_{n_-}}\no\\
    \simeq&\,
    \braket{P_+^x}\braket{P_+^x\cO P_+^x}_{\rm mc}^{\delta E_+}(\cE_+)
    + \braket{P_-^x}\braket{P_-^x\cO P_-^x}_{\rm mc}^{\delta E_-}(\cE_-),
    \label{stationary}
\end{align}
where $\ket{E^+_{n_+}}$ and $\ket{E^-_{n_-}}$ denote the eigenstates with $U(C_x^*)=+1$ and $U(C_x^*)=-1$, respectively.
Here, we have used the ETH for each symmetry sector 
\begin{align}
\braket{E^\pm_{n_\pm}|\cO|E^\pm_{n_\pm}}\simeq \braket{P_\pm^x\cO P_\pm^x}_{\rm mc}^{\delta E_\pm}(E_{n_\pm}^\pm)
\end{align}
and assumed that $\frac{|c_{n_\pm}|^2}{\sum_{n_\pm}|c_{n_\pm}|^2}$ is localized around the mean energy for each symmetry sector
\begin{align}
\cE_\pm=\frac{\sum_{n_\pm}|c_{n_\pm}|^2E_{n_\pm}}{\sum_{n_\pm}|c_{n_\pm}|^2}=\frac{\braket{P_{\pm}^x H_{\mathbb{Z}_2}P_{\pm}^x}}{\braket{P_{\pm}^x}},
\end{align}
where $\sum_{n_\pm}$ stands for  $\sum_{{n_\pm}:U(C_x^*)=\pm 1\text{ sector}}$ in the following.
 Note that the precise values of $\delta E_\pm$ are not important, and we will not consider them in the following.

Next, we consider the GGE. For this purpose, we define canonical ensembles for each sector
 \begin{align}\begin{split}
     \braket{\cO}_{{\rm can},\pm} :=
     \frac{1}{Z_{\pm}}\sum_{n_{\pm}} \cO_{n_{\pm}n_{\pm}}
     e^{-(\beta\pm \mu_x) E^{\pm}_{n_{\pm}} \mp \lambda_x},\qquad
     Z_{\pm}:=\sum_{n_{\pm}}e^{-(\beta\pm \mu_x) E^{\pm}_{n_{\pm}} \mp \lambda_x}
     .
 \end{split}\end{align}
 In terms of these ensembles, the GGE (\ref{sup:GGE-definition}) is expressed as
\begin{align}
    \operatorname{Tr}(\cO \rho_{\rm GGE} ) =&\,
    \frac{1}{Z_{\rm GGE}} 
    \bigg( 
    \sum_{n_+}e^{-\beta E_{n_+} -\lambda_x - \mu_x E_{n_+}}\cO_{n_+n_+}
    + \sum_{n_-}e^{-\beta E_{n_-} +\lambda_x + \mu_x E_{n_-}}\cO_{n_-n_-}
    \bigg)\no\\
    =&\,
    \sum_{s=\pm}\frac{Z_{s}}{Z_{\rm GGE}} \braket{\cO}_{{\rm can},s}
    = \sum_{s=\pm}\braket{P^x_{s}}_{\rm GGE}\braket{\cO}_{{\rm can},s}.
\end{align}
As for the standard canonical ensemble, by appropriately choosing $\beta\pm\mu_x$, we can set
\begin{align}\label{mubeta}
\cE_{\pm}=\braket{H}_{{\rm can},\pm}=\frac{\braket{P_{\pm}^x H_{\mathbb{Z}_2}P_{\pm}^x}_\mathrm{GGE}}{\braket{P_{\pm}^x}_\mathrm{GGE}}.
\end{align}
Under this condition,
we assume that the distributions $\braket{E^{\pm}_{n_{\pm}}|\rho_{\pm}|E^{\pm}_{n_{\pm}}}= \exp(-(\beta\pm \mu_x) E^{\pm}_{n_{\pm}} \mp \lambda_x)/Z_{\pm}$ are sufficiently localized around $\cE_{\pm}$. Then, using the ETH for each symmetry sector, we can write the operator averages as
\begin{align}
    \braket{\cO}_{{\rm can},\pm} \simeq
    \frac{1}{Z_{\pm}}\sum_{n_{\pm}} 
     e^{-(\beta\pm \mu_x) E^{\pm}_{n_{\pm}} \mp \lambda_x} \braket{P_{\pm}^x\cO P_{\pm}^x}_{\rm mc}(\cE_{\pm})
     = \braket{P_{\pm}^x\cO P_{\pm}^x}_{\rm mc}(\cE_{\pm}).
\end{align}
Choosing $\lambda_x$ appropriately, we can set
\begin{align}\label{lambda}
\braket{P^x_{+}}_{\rm GGE}= \braket{P_{+}^x}, \qquad
\big(\Rightarrow \quad \braket{P^x_{-}}_{\rm GGE} = 1 - \braket{P^x_{+}}_{\rm GGE} = 1 - \braket{P_{+}^x}
=\braket{P_{-}^x} \big).
\end{align}

In summary, by setting $\beta,\mu_x$, and $\lambda_x$ such that Eqs.~\eqref{mubeta} and~\eqref{lambda} hold true, we can show that
the long-time average of the observable $\cO$ (\ref{stationary}) can be identified with the GGE prediction:
\begin{align}
    \overline{\braket{\psi|\cO|\psi}} \simeq&\,
    \sum_{\pm}\braket{P^x_{\pm}}_{\rm GGE}\braket{\cO}_{{\rm can},\pm}
    =\operatorname{Tr}(\cO \rho_{\rm GGE} ).
\end{align}

Note that the conditions Eqs.~\eqref{mubeta} and~\eqref{lambda} are equivalent to the following condition for the conserved quantities in the GGE,
\begin{align}\label{GGEcondition}
\braket{H_{\mathbb{Z}_2}}_\mathrm{GGE}=\braket{H_{\mathbb{Z}_2}},\:\braket{U(C_x^*)}_\mathrm{GGE}=\braket{U(C_x^*)},\:\braket{U(C_x^*)H_{\mathbb{Z}_2}}_\mathrm{GGE}=\braket{U(C_x^*)H_{\mathbb{Z}_2}},
\end{align}
as expected.

The above discussion has  focused on observables that are non-local in the $x$-direction but local in the $y$-direction.
To adopt operators extended to the $y$-direction as well, we should include at most seven chemical potentials in total as
$
\tilde{\rho}_{\rm GGE} = \tilde{Z}_{\rm GGE}^{-1}
\exp\big(-\beta H_{\mathbb{Z}_2} 
- \sum_{i=x,y} \lambda_i U(C^*_i) 
-\sum_{i=x,y}\mu_i U(C^*_i) H_{\mathbb{Z}_2}
-\alpha U(C_x^*)U(C_y^*) -\alpha' U(C_x^*)U(C_y^*) H_{\mathbb{Z}_2}
\big)
$.

\subsection{Justification for the general case}

In the case of a general finite abelian group $G$, we propose that  the following density matrix serves as the GGE:
\begin{align}
\rho_\mathrm{GGE}^G(\beta^G,\{\lambda^G_j\},\{\gamma_j^G\}) =\frac{1}{Z_\mathrm{GGE}^G(\beta^G,\{\lambda^G_j\},\{\gamma_j^G\}) }
e^{-\beta^G H - \sum_{j=1}^{N-1} \lambda^G_j P_j - \sum_{j=1}^{N-1} \mu_j^G P_j H},
\label{GGE-general}
\end{align}
where $P_j$ $(j=1,\dots,N)$ are the projections to each symmetry sector, and the number of the sectors is given by $N=|H_{d-p}(\cM,G)|$.
Note here that the summation over the chemical potentials is performed over $1\leq j \leq N-1$ because operators $P_N$ and $P_N H$ are not independent of the other conserved quantities.

Assuming the ETH for each symmetry sector, 
\begin{align}
\braket{E_{n_j}^j| \cO|E_{n_j}^j}
=\braket{P_j \cO P_j }_{\rm mc}^{\delta E_j}(E_{n_j}^j)\quad(1\leq j\leq N),
\end{align}
we can obtain the stationary state as in the $\mathbb{Z}_2$ case (\ref{stationary}):
\begin{align}
 \overline{\braket{\psi|\cO|\psi}} =&\,
    \sum_{n}|c_n|^2 \braket{E_n|\cO|E_n}
    =
    \sum_{j=1}^{N}\sum_{n_j}|c_{n_j}|^2 
    \braket{E_{n_j}^j| \cO|E_{n_j}^j}\no\\
    \simeq&\,\sum_{j=1}^N
    \braket{P_j}\braket{P_j \cO P_j }_{\rm mc}^{\delta E_j}(\cE_j),
    \label{general-time}
\end{align}
where $|E_{n_j}^j\rangle$ are energy eigenstates in each symmetry sector and 
\begin{align}
\cE_j=\frac{\braket{P_jHP_j}}{\braket{P_j}}.
\end{align}
Here, $\sum_{n_j}$ means that the sum is taken only over eigenstates that belong to the symmetry sector $j$.
We have also assumed the localization of the energy distribution in each sector around $\cE_j$.
In the following, we omit $\delta E_j$, whose details are not important in the thermodynamic limit.

Along the same lines as the $\mathbb{Z}_2$ case, the GGE (\ref{GGE-general}) can be written as
\begin{align}
 \braket{\cO}_{\rm GGE}^G=&\,
 \frac{1}{Z_{\rm GGE}^G} 
 \left(\sum_{j=1}^{N-1}\sum_{n_j}e^{-(\beta^G+\mu_j^G) E_{n_j} - \lambda_j }\cO_{n_jn_j}
 + \sum_{n_N}e^{-\beta^G E_{n_N} }\cO_{n_Nn_N}
 \right)\no\\
 =&\,
 \sum_{j=1}^N \frac{Z_{j}}{Z_{\rm GGE}^G} \braket{\cO}_{{\rm can},j}
 = \sum_{j=1}^N \braket{P_j}_{\rm GGE}^G \braket{\cO}_{{\rm can},j},
\end{align}
where canonical ensembles restricted to each sector are defined by
\begin{align}\begin{split}
\braket{\cO}_{{\rm can},j}:=&\, \frac{1}{Z_j}\sum_{n_j}\cO_{n_jn_j}e^{-(\beta^G + \mu_{j}^G) E_{n_j} - \lambda_j^G },
\qquad
Z_{j} := \sum_{n_j} e^{-(\beta^G + \mu_{j}^G) E_{n_j} - \lambda_j^G },
\qquad (j=1,\dots,N-1),\\
\braket{\cO}_{{\rm can},N}:=&\, \frac{1}{Z_N}\sum_{n_N}\cO_{n_Nn_N}e^{-\beta^G E_{n_N} },
\qquad
Z_{N} := \sum_{n_N} e^{-\beta^G E_{n_N}  }.
\end{split}\end{align}

Here, we can choose $\beta^G$ and $\mu_j^G$ ($j=1,\cdots,N-1$) such that
\begin{align}\label{Ej}
\cE_j=\braket{H}_{\mathrm{can}, j}=\frac{\braket{P_jHP_j}_\mathrm{GGE}^G}{\braket{P_j}^G_\mathrm{GGE}}
\end{align}
holds for $1\leq j\leq N$.
Now, the assumption that the distributions $\braket{E_{n_j}|\rho_j|E_{n_j}}:=\exp(-(\beta^G + \mu_{j}^G) E_{n_j} - \lambda_j^G )/Z_j$ and $\braket{E_{n_N}|\rho_{N}|E_{n_N}}:=\exp(-\beta^G E_{n_N})$ are localized around $\cE_j$ again leads to the following:
\begin{align}
\braket{\cO}_{\mathrm{can},j} = \braket{P_j \cO P_j}_{\rm mc}(\cE_j),\qquad
(j=1,\dots,N).
\end{align}
We can determine $\lambda_j^G\:(j=1,\dots,N-1)$ uniquely such that the following holds true:
\begin{align}\label{Pj}
    &\braket{P_j}_{\rm GGE}^G= \braket{P_j},\qquad (j=1,\dots,N-1),\\
    \Rightarrow \quad &
    \braket{P_N}^G_{\rm GGE}=\frac{Z_N}{Z_{\rm GGE}} = 1 - \sum_{j=1}^{N-1}\frac{Z_j}{Z_N}
    = 1-\sum_{j=1}^{N-1}\braket{P_j} = \braket{P_N}.
\end{align}

In summary, by choosing $\beta^G, \lambda_j^G,$ and $\mu_j^G\:(j=1,\cdots,N-1)$ such that Eqs.~\eqref{Ej} and~\eqref{Pj} hold true, we can show that
the time average (\ref{general-time}) is given by
\begin{align}
\overline{\braket{\psi | \cO | \psi}}=&\,
\sum_{j=1}^N\braket{P_j}_{\rm GGE}^G \braket{\cO}_{\mathrm{can},j}
= \braket{\cO}_{\rm GGE}^G .
\end{align}
Finally, the conditions Eqs.~\eqref{Ej} and~\eqref{Pj} are equivalent to the following condition for the conserved quantities in the GGE,
\begin{align}
\braket{H}_\mathrm{GGE}^G=\braket{H},\:\braket{P_j}^G_\mathrm{GGE}=\braket{P_j},\:\braket{P_jH}^G_\mathrm{GGE}=\braket{P_jH}\:\:(1\leq j\leq N-1),
\end{align}
where we have used $P_jHP_j=P_jH$ and $\braket{H}=\sum_{j=1}^N\braket{P_jHP_j}$.

Recalling the relation (\ref{projection}), we can recover the GGE (\ref{sup:GGE-definition}) by redefining $\beta^G$, $\lambda_j^G$, and $\mu_j^G$ for the $G=\mathbb{Z}_2$ case.
In addition, similar reasoning justifies the GGE $\tilde{\rho}_\mathrm{GGE}$ for $G=\mathbb{Z}_2$ discussed in the main text, where we consider the non-local observables both for the $x$- and $y$-directions.


%